
\input harvmac
\sequentialequations
\noblackbox
\def\tit{\bigskip\centerline{\it Department of Physics}
\centerline{\it Tokyo Institute of Technology}
\centerline{\it Oh-okayama, Meguro-ku}
\centerline{\it Tokyo, JAPAN  152} \vskip .3in}
\def\lesssim{\mathrel{\raise.3ex\hbox{$<$\kern-.75em\lower1ex\hbox{$\sim$}}}}
\def\nf{N_F}
\def\nc{N_C}
\def\al#1{\alpha_{#1}}
\def\drhotone{\delta\rho\bigr|_{top,1-loop}}
\def\drhohone{\delta\rho\bigr|_{Higgs,1-loop}}
\def\drhot{\delta\rho\bigr|_{top}}
\def\drhoh{\delta\rho\bigr|_{Higgs}}
\def\gf{G_F}
\def\mt{m_t}
\def\ts{\tilde S}
\def\yt{y_t}
\def\gp{g^{\prime2}}
\def\nl{\hfil\break}
\def\lnl{\log(\Lambda^2/\strivh)}

\def\lh{\lambda_{{}_H}}
\def\any{A_n^{y}}
\def\anl{A_n^{\lambda}}
\def\coupy{\al y}
\def\coupl{\al\lambda}
\def\O{{\cal O}}
\def\hmt{m_{t}}
\def\hmh{M_H}
\def\mh{M_H}
\def\thw{\theta_W}

\def\npb{{ \sl Nucl. Phys. }}

\def\prd{{ \sl Phys. Rev. }}
\def\prl{{ \sl Phys. Rev. Lett. }}
\def\plb{{ \sl Phys. Lett. }}
\def\rmp{{ \sl Rev. Mod. Phys. }}
\def\strivt{s_{triv}^t}
\def\strivh{s_{triv}^H}

\def\szero{s_0}
\def\mw{M_W}
\def\mz{M_Z}
\def\intkl{\int_{k^2<\Lambda^2}\!\! {d^4k\over(2\pi)^4}}
\def\intz{\int_0^\infty\!\!\!\!dz\,}
\def\intx{\int_0^\infty\!\!\!\!dx\,}

\nopagenumbers\abstractfont\hsize=\hstitle\rightline{TIT/HEP--235}
\vskip 1in\centerline{\titlefont
Large--order behavior of non--decoupling effects and triviality
} \abstractfont\vskip .5in\pageno=0
\centerline{Kenichiro Aoki\footnote{$^\dagger$}{%
\hskip-1mm email:{\tt~ken@phys.titech.ac.jp}}}
\baselineskip=11.5pt plus 2pt minus 1pt
\tit
\bigskip
\centerline{\bf Abstract}
We compute some non--decoupling effects in the standard
model, such as the $\rho$ parameter, to all orders
in the coupling constant expansion.
We analyze their large order behavior and explicitly
show how it is related to the non--perturbative
cutoff dependence of these non--decoupling effects
due to the triviality of the theory.

\Date{9/93}
\newsec{Introduction}
Non--decoupling effects, such as the $\rho$ parameter,
 have played  a crucial role in restricting the parameters
of the  standard model using the
precision electroweak measurements
\ref\RADIATIVE{For a review on
radiative corrections in the standard model, see for instance,
K.~Aoki, Z.~Hioki, R.~Kawabe, M.~Konuma, T.~Muta, {\sl Suppl.
Progr. Theor. Phys.,} {\bf 73} (1982) 1\nl
For a recent comparison with experiments, see for instance,
P.~Langacker, M.~Luo, A.~Mann, \rmp{\bf64} (1992) 87
}.
In these effects, the quantum effects of heavy particles
in low energy physical observables are not suppressed by
the heavy particle mass.
In contrast to the cases where the decoupling theorem
applies, these cases arise when the increase in the
particle mass is accompanied
by an increase in the strength of interactions, as it is in the
standard model \ref\AC{T.~Appelquist,
J.~Carazzone,  \prd{\bf11} (1975) 2856 \nl J.C.~Collins,
{\sl ``Renormalization"}, Cambridge University Press (1984),
sect. 8}.

In another direction, non--perturbatively, it has been established
that for weak gauge couplings, the standard
model is a theory defined with a physical cutoff at high energies,
namely the triviality scale
\ref\TRIVOLD{L.D.~Landau, I. Pomeranchuk, {\sl Dokl. Akad. Nauk.
USSR}, {\bf102} (1955) 489\nl
K.G. Wilson, \prd{\bf D7} (1973) 2911\nl
M. Aizenman, \prl{\bf47} (1981) 1\nl
J. Fr\"ohlich, \npb{\bf B200} (1982) 281\nl
W.A.~Bardeen, M. Moshe, \prd{\bf D28} (1983) 1373\nl
M.B.~Einhorn, \npb{\bf B246} (1984) 75\nl
M.~Luscher, P.~Weisz, \npb{\bf B290} (1987) 25,
{\bf B295} (1988) 65, \plb{\bf 212B} (1988) 472}\ref\TRIVIALITY{
A.~Hasenfratz, K.~Jansen, C.B.~Lang, T.~Neuhaus, H.~Yoneyama, \plb{\bf 199B}
(1988) 531\nl
J.~Kuti, L.~Lin, Y.~Shen, \prl{\bf61} (1988) 678\nl
J.~Jers\'ak, in {\sl``Eloisatron Project Workshop",}  (1988)}%
\ref\NFERM{M.B.~Einhorn, G.~Goldberg, \prl{\bf57 }(1986) 2115\nl
K. Aoki, \prd{\bf D44} (1991) 1547}%
\ref\FLAT{J.~Shigemitsu, \plb{\bf226B} (1989) 364\nl
I-H.~Lee, J.~Shigemitsu, R.~Shrock, \npb{\bf B330} (1990) 225,
\npb{\bf B335} (1990) 265\nl
W.~Bock, A.K.~De, C.~Frick, K.~Jansen, T.~Trappenburg,
\npb{\bf B371}~(1992)~683\nl
S.~Aoki, J.~Shigemitsu, J.~Sloan, \npb{\bf B372} (1992) 361
}.
This scale is always above the symmetry breaking scale
for consistency and becomes lower as the top (or the Higgs)
becomes heavier.
The quantities measured at low energies should be insensitive
to these high energy effects if the decoupling theorem applies.
In non--decoupling effects, this does not necessarily hold since it
actually sees the physics of the symmetry breaking scale
even though they are measured at low energies.
Indeed, it has been shown non--perturbatively using the $1/\nf$ expansion
that when the cutoff is of the order of the symmetry breaking scale,
the non--decoupling effects inevitably exhibit non--universal behavior
\ref\AP{K. Aoki, S. Peris, UCLA preprint, UCLA/92/TEP/23 (1992)}\ref\APS{
K. Aoki, S. Peris, \prl{\bf 70} (1993) 1743}.
(See also  \ref\CPP{S. Cortese,
E. Pallante, R. Petronzio, \plb{\bf 301B} (1993) 203}
where a similar conclusion is drawn
from another approach.)
This ambiguity is an inherent limitation of the standard model
which needs to be defined with a high energy cutoff.
The cutoff dependence of non--decoupling effects is the
sensitivity at low energies
to the physics beyond the standard model.

In this work, we will explicitly compute
the perturbative behavior of some non--decoupling
effects in the standard model to all orders,
to leading order in the $1/\nf$ expansion.
Then we will show, in some cases, how the large order behavior
is related to the aforementioned cutoff dependence of
these non--decoupling effects  necessitated by the
triviality of the standard model.
There has been considerable amount of work on the
large order behavior of perturbation theory. (For a
general review and references on the subject, see,
for instance \ref\THT{G. {}'t Hooft, lectures given at
Erice (1977)}\ref\LO{J.C.~Le~Guillou, J.~Zinn-Justin, (eds.),
{\sl``Large order perturbation theory"}, North Holland (1990)}.)
To our knowledge, our work is novel in that we explicitly show
how the large order behavior of perturbation theory is
related to the physical effects of triviality
using a controlled non--perturbative approximation.
In non--decoupling effects, these non--perturbative effects become
substantial when the top or the Higgs mass is large.

First, let us consider the top contribution to the $\rho$
parameter in the standard model. The one loop expression,
to leading order in the gauge couplings, is
\eqn\rhotoneloop{
  \drhotone={3\over(4\pi)^2}\sqrt2\gf\mt^2
  = {3 \yt^2\over 32\pi^2}
}
Here, $\yt$ is the Yukawa coupling for the top and $\gf$ is the
Fermi's constant  $(\sqrt2\gf)^{-1}\equiv v^2=(246\,GeV)^2$.
We see that this is a typical non--decoupling
effect wherein the contribution
to a low energy measurable parameter from a virtual top
increases as a power of the top mass rather than fall off.
The top contribution to the $\rho$ parameter to leading order
in the $1/\nf$ expansion is \AP
\eqn\rhotnonpert{
  \drhot={\nc\over v^2}\intkl{\hmt^4\over k^2\left[
       \left(1-\coupy\log(k^2/\szero)\right)k^2+\hmt^2\right]^2}}
$\nf$ is the number of flavors and $\nc$ is the number of
colors; in the standard model, $\nf=2$ and $\nc=3$.
The  Yukawa coupling constant, $\yt$, renormalized at the scale $\szero$
has been used in the expression and the notation $\coupy\equiv
{\yt^2\nf/(32\pi^2)}$ has been used for brevity.
Also, $\hmt^2\equiv\yt^2v^2/2$.
The renormalization point $\szero$ is arbitrary.
The integral is {\it not} finite unless we cutoff the integral
at the scale $\Lambda^2$. The cutoff is $\Lambda^2$ is smaller
than, but is of the same order as the triviality scale,
$\strivt=\szero\exp(1/\coupy)$.
This triviality scale has a typical non--perturbative dependence
on the coupling and is a physical parameter independent of
the renormalization scale $\szero$.
We note that the cutoff for the non--perturbative expression
for the $\rho$ parameter in \rhotnonpert\ is not only
natural from the point of view of triviality, but {\it
necessary}.
The existence of the cutoff leads to a cutoff dependence of the
$\rho$ parameter.
These non--universal effects can be substantial when the cutoff is
of the order of the symmetry breaking scale, $v^2$.

Let us analyze $\drhot$ in the perturbative context:
To leading order in the $1/\nf$ expansion, we may obtain
the perturbative series to all orders in the coupling
constant expansion as
\eqn\rhotpert{
  \drhot={1\over\nf}\sum_{n=0}^\infty\any\coupy^{n+1},\qquad
     \any\equiv(n+1)\intx{(x\log x)^n\over(x+1)^{n+2}}}
In the perturbative context, there is no triviality scale;
it is both unnatural and unnecessary to cutoff the momentum integrals
and we shall not do so.
We have chosen to renormalize at the scale $\szero=\hmt^2$
for convenience.

The large order behavior of the coefficients of the
perturbative series may be obtained using a
saddle point approximation as
\eqn\anysaddle{
  \any=(n+1)!\left(1+\O({1\over n^2})\right)}
Numerically, the agreement is better than  0.1\%
for $n\geq14$.
A couple of comments are in order:
First, we see that the perturbative series has zero
radius of convergence.
This is not surprising since we do not expect the theory
to make sense when the Yukawa coupling constant
is imaginary \ref\DYSON{F.J. Dyson, \prd{\bf 85} (1952) 631}.
Second, the perturbative series is not even Borel summable\ref\WW{
E.T. Whittacker, G.N. Watson, {\sl ``A course in modern analysis''},
Cambridge University Press (1902)}.
That is, using the integral expression for $n!$, we may
derive an integral expression incorporating the higher
order behavior of the  perturbative
series for $\drhot$ as
\eqn\drhotborel{
  \drhot\simeq{1\over\coupy\nf}\intz{z\,e^{-z/\coupy}\over1-z}}
This expression is ill--defined due to the existence of a
pole on the positive real axis in the integrand.
{}From this,  we see that there is no obvious way to obtain
an unambiguous prediction from the perturbative result.
Within the perturbative context, it is not clear whether this
is a limitation of the perturbation theory or something deeper.

In view of the non--perturbative understanding explained
previously, we may understand the physics underlying
the breakdown of the perturbation theory in $\drhot$.
Using the resummed expression \drhotborel, we could
try to make sense out of the whole perturbative
series by somehow regulating the integral, for instance,
by cutting off the integral at $z<1$.
This, however, leads to inherent ambiguities in the
result, which we can estimate to be of order $1/\coupy\exp(-1/\coupy)$.
These ambiguities are none other than the cutoff dependence
of the non--perturbative result for $\drhot$ in \rhotnonpert\ introduced
by triviality. Indeed, these cutoff effects may be obtained from
\rhotnonpert\ to be $\O(v^2/\strivt)=\O(1/\coupy\exp(-1/\coupy))$.
As previously mentioned, the renormalization scale $\szero$
was chosen to be at $\hmt^2$.
Had we chosen another scale, the coupling constant
expansion would have been slightly
more complicated, but it would not have changed the
properties of the perturbative series, in particular, its non
Borel summability and the consequent ambiguity.
This is consistent with the fact that this ambiguity, or equivalently,
the cutoff effects in $\drhot$ are physical so that it should
not depend on the choice of the renormalization scheme at all.

Next, we consider the Higgs contribution to the $\rho$ parameter
in the standard model.
The one loop expression is
\eqn\rhohoneloop{
  \drhohone=-{3\over4}{g^{\prime2}\over(4\pi)^2}\log{\mh^2\over\mw^2}}
In this case, the low energy observable grows logarithmically
with respect to the heavy particle mass,
in accordance with the screening theorem \ref\SCREENING{M.~Veltman,
{\sl Acta Phys. Pol.}{ \bf B8}
(1977) \plb{\bf 70B}, (1977) 253\nl M.B.~Einhorn, J.~Wudka,
\prd{\bf 39D} (1989) 39}.
The non--perturbative expression for $\drhoh$, to leading
order in the $1/\nf$ expansion, is \AP
\eqn\drhohnonpert{
  \drhoh=-{3\over4}\gp\intkl{\hmh^2\over(k^2+\mw^2)(k^2+\mz^2)
\left[(1-\coupl\log k^2/\szero)k^2+\hmh^2\right]}}
Here, we used the notations $
\mw\equiv gv/2,\ \mz=gv/(2\cos\thw),\
\hmh^2\equiv2\lh v^2$, $\coupl\equiv\lh\nf/(8\pi^2)$ and
$\lh$ is the renormalized  coupling constant at the scale $\szero$.
As in the top case, the non--perturbative expression is not
finite unless a cutoff $\Lambda^2$ is imposed, which leads
to the cutoff dependence of the result.
The triviality scale is $\strivh\equiv
\szero\exp(1/\coupl)$ and the cutoff is $\Lambda^2\lesssim\strivh$.

We may obtain the perturbative series to all orders
to leading order in the $1/\nf$ expansion as
\eqn\rhohpert{
  \drhoh=-{3\over4}{\gp\over(4\pi)^2}\sum_{n=0}^\infty\anl\coupl^n,\qquad
    \anl\equiv\intx{x^{n+1}\log^nx\over
       (x+\beta_1)(x+\beta_2)(x+1)^{n+1}}}
where $\beta_1\equiv\mw^2/\hmh^2,\ \beta_2\equiv\mz^2/\hmh^2$.
The large order behavior of the coefficients in the series is
\eqn\anlsaddle{
  \anl=n!\left(1+\O({1\over n^2})\right)}
This is similar to the top case and the resummed series
is
\eqn\drhohborel{
  \drhoh\simeq\left(-{3\over4}g'^2\right)
  {1\over\coupl}\intz{e^{-z/\coupy}\over1-z}}
which again is ill--defined due to the pole at $z=1$.
This leads to the ambiguity in the resummed series of order
$g'^2v^2/\strivh$ which again may be identified with the cutoff
effects in $\drhoh$ due to triviality.
As we can see from the one loop results \rhotoneloop,\rhohoneloop,
the top and the Higgs cases differ qualitatively.
However, the non--perturbative ambiguity due to triviality
and its relation to the large order behavior of their perturbative
expansion is essentially the same, apart from the overall factor
of $g'^2$.

The singularity in the Borel transform we have found
is ultraviolet in origin and
is sometimes called a ``renormalon'' effect
\THT\ref\RENORMALON{D.J. Gross, A. Neveu, \prd{\bf D10} (1974) 3235\nl
B. Lautrup, \plb{\bf 69B} (1977) 438\nl
S. Chadha, P. Olesen, \plb{\bf 72B} (1977) 87\nl
G. Parisi, \plb{\bf 73B} (1978) 65
}.
This should be distinguished from  the divergence
of perturbation series which is infrared in origin, such
as the behavior that is associated with the classical
solutions in the massless scalar theory \ref\LIP{L.N. Lipatov,
{\sl JETP Lett. }{\bf 25} (1977) 104, {\sl Sov. Phys. JETP }{\bf
45} (1977) 216}.
In the theories which have no asymptotic freedom,
this second kind of divergence result in
alternating signs for the coefficient of the
perturbation series, which can produce
singularities on the negative real axis in the Borel transform.
This will not prevent us from resumming the perturbation theory
to obtain an unambiguous number.
The role of the infrared and the ultraviolet are thought to be
reversed in asymptotically free theories.

To summarize, the perturbative series for $\drhot,\drhoh$ are not
only divergent but also not Borel summable.
This leads to ambiguities in the resummed series for $\drhot,\drhoh$
which can be identified with the non--universal effects in these
non--decoupling effects introduced
due to triviality of the standard model.
It is important to note that this argument cannot be reversed.
It is possible for additional non--perturbative contributions
to exist that is not apparent within the perturbation theory,
even if we compute the perturbative series to all orders.
In particular, even if the series is convergent, there
could be non--perturbative contributions.
This is clearly illustrated in the next example.

Let us consider the heavy fermion
contribution to the so--called $S$ parameter
\ref\SREF{
M.~Peskin, T.~Takeuchi, \prl{\bf 65} (1990) 964
}
from the longitudinal modes of the gauge bosons.
This parameter which we call $\ts$, is to one loop,
\eqn\tsoneloop{
   \ts={\nc\over12\pi}}
per one heavy fermion multiplet.
In this non--decoupling effect, the contribution from
the heavy particles is a constant with respect to
the heavy particle mass.
To leading order in the $1/\nf$ expansion, the non--perturbative
expression for this parameter is \APS
\eqn\tsnonpert{
  \ts={\nc\over12\pi}
  \left[1+\gamma^2{\left(2\lnl+3\right)
   +4\gamma\left(\lnl+1\right)-\gamma^2\over
  (-\lnl+1)^4}\right]\quad\hbox{where }\gamma\equiv{(4\pi)^2\over\nf}
  {v^2\over\Lambda^2}}
The perturbative expansion for this parameter is exactly
\tsoneloop\ to {\it all orders.}
The perturbative series is not only Borel summable, but it
has an infinite radius of convergence.
However this observable also has the same kind of cutoff effects
as $\drhot,\drhoh$.
The cutoff dependence of $\ts$ is of  $\O(v^4/(\strivh)^2)=
\O(\coupy^{-2}\exp(-1/(2\coupy)))$,
which is completely hidden from perturbation theory.
It should perhaps be pointed out, however, that in the
same theory, there also exist non--decoupling effects
whose perturbation series are non--Borel summable,
such as $\drhot$ in \rhotpert.

In closing, we would like to comment on non--decoupling
effects in supersymmetric theories.
In general, the quantum properties of supersymmetric theories
qualitatively differ from those of the non--supersymmetric
theories, as exemplified by the so--called ``non--renormalization
theorems''.
It has been shown, however, that the supersymmetric standard
model is also a theory defined with a cutoff, namely the triviality
scale, in a manner similar to the standard model
\ref\KASUSY{K. Aoki, \prd{\bf D46} (1992) 1123}.
The $\rho$ parameter has been computed non--perturbatively
in some cases and it has cutoff dependence due to triviality
as in the non--supersymmetric case \ref\KASRHO{K. Aoki,
Tokyo Institute of Technology preprint, TIT/HEP--233}.
The contribution of the top supermultiplet to the
$\rho$ parameter is essentially the same as $\drhot$
analyzed above when there are no soft breaking terms and
no mixing of the Higgs supermultiplets.
Therefore, at least in this case, the relation between
the large order behavior of perturbation theory and the
non--universal effects due to triviality may be understood
in a manner identical to the non--supersymmetric case.
We expect the above understanding of the relation between
the perturbation theory and the non--perturbative
effects due to triviality to apply in general to non--decoupling effects in
supersymmetric theories with triviality.
\medskip\noindent
{\bf Acknowledgments:} We would like to thank Norisuke Sakai
and Hidenori Sonoda for encouragement and discussions.
\listrefs
\end